\documentclass[aps,prb,superscriptaddress,twocolumn,showpacs]{revtex4}

\usepackage{amsmath,graphicx,color}
\begin{document}
\title{The exact solution of the diffusion trapping model of defect profiling with variable energy positrons}
\author{V.A.Stephanovich}\email{stef@uni.opole.pl}
\address{Institute of Physics, Opole University, ul. Oleska 48, 45-052 Opole, Poland}
\author{J. Dryzek} \email{jdryzek@gmail.com}
\address{Institute of Physics, Opole University, ul. Oleska 48, 45-052 Opole, Poland}
\address{Institute of Nuclear Physics PAN, ul. Radzikowskiego 152, 31-342 Krak\'ow, Poland}
\date{\today}

\begin{abstract}
We report an exact analytical solution of so-called positron diffusion trapping model.
This model have been widely used for the treatment of the experimental data for defect profiling of the adjoin surface layer using the variable energy positron (VEP) beam technique. Hovewer, up to now this model could be treated only numerically with so-called VEPFIT program. The explicit form of the solutions is obtained for the realistic cases when defect profile is described by a discreet step-like function and continuous exponential-like function. Our solutions allow to derive the analytical expressions for typical positron annihilation characteristics including the positron lifetime spectrum. Latter quantity could be measured using the pulsed, slow positron beam. Our analytical results are in good coincidence with both the VEPFIT numerics and experimental data. The presented solutions are easily generalizable for defect profiles of other shapes and can be well used for much more precise treatment of above experimental data.
\end{abstract}
\pacs{78.70.Bj, 41.75.Fr, 68.35.Dv, 61.72.J-}
\maketitle

\section{Introduction}

Defect depth profiling near the surface is an important application of the positron annihilation spectroscopy. For this purpose, the conventional techniques based on the isotope positron source have been successfully applied, in which case the defect distribution extended up to the depth of hundreds micrometers is possible to measure \cite{dr13}. The detection of the near-surface defect distribution (up to the micrometer or less depths) is also possible, however with a slow positron beam. This technique, in particular, is suitable to study the metals and semiconductors with their surfaces being implanted with different ions, \cite{p2,p3}. For efficient depth profiling, the positron data analysis should be extended beyond the commonly used simple two-state trapping model. Even in the above conventional techniques positron implantation profile should be taken into account \cite{dr13}. This is because emitted energetic positrons exhibit the energy distribution which finally leads to a positron implantation profile. It can happen, that the total depth of the defects distribution correlates with that for implantation profile. For the efficient treatment of the data obtained from the slow positron beam measurements, not only the initial implantation profile, but also the thermalized positrons diffusion should be considered. This is taken into account in the positron diffusion trapping model (DTM).

The DTM is an extended version of the trapping model, where only annihilation and trapping rates are introduced. Namely, they are included into the set of rate equations which describe the time evolution of positron populations in different states. No positron dynamics is considered in the trapping model. On the contrary, the DTM accounts for the positrons thermal motions prior to annihilation and after implantation. The epithermal motion is also possible to introduce. These motions are represented in DTM as certain diffusion processes. In this case, the implantation profile is included as an initial condition for the corresponding time-dependent diffusion equation. The solution of the rate equations is necessary to obtain the time evolution of the positron fractions at different states. This permits to deduce the measurable positron annihilation characteristics (PACh) like the lineshape S - parameter of the Doppler-broadened annihilation line, mean positron lifetime and positron lifetime spectrum.  

The development of the variable and pulsed slow positron beam techniques, which allows measuring the positron lifetime spectrum nearby and at the surface require the solution of DTM equations. Several approaches have been applied for this purpose. The one dimensional diffusion equation including both annihilation in the sample interior (the bulk annihilation) and the surface trapping have been considered by Frieze et al.\cite{p4}. To solve the corresponding diffusion equation, the authors \cite{p4} used Fourier transformation technique.  Britton \cite{p5}, used Green's function method for solution of the diffusion equation assuming the perfect homogeneous defects-free solid. In Ref. \cite{p5}, the approximate forms for positron populations at the surface and in the bulk have been derived. The final results have been obtained numerically. The author \cite{p5} have also considered the effects of epithermal positrons reaching the surface as well as internal reflection of thermal positrons. K\"ogel \cite{p6} has elaborated the DTM in a solid with inhomogeneous defects distribution. However, the explicit solutions in Ref. \cite{p6} have been obtained only for the standard trapping model with homogeneous defect distribution. The exact solution of the DTM for homogeneous, defect free  sample as well as that with uniformly distributed open volume defects was made possible using time domain Laplace transformation and a Green's function method to solve resulting coordinate equation \cite{p7}. In this case, the PACh were expressed in the closed analytical form as the functions of the DTM parameters. It is interesting to note, that the solution \cite{p7} predicts that the positron lifetime spectrum in a pulsed beam experiment cannot be expressed as a simple sum of exponential functions similar to the standard trapping model. We note also that the pulsed beam technique, which allows  to obtain the positron lifetime spectra as the function of its energy, is not completely operational due to the design difficulties.

The beams, where only the positron energy can be varied, i.e., the variable energy positron beam (VEP), are much more popular. In this case,  the annihilation lineshape parameter is measured as a function of positron energy. To solve the DTM for that case, only the steady-state diffusion equations solution is necessary. In this solution, the corresponding time dependence is integrated from zero to infinity. This case is much simpler as it does not require the inversion of Laplace transform and allows considering the inhomogeneous case where defect concentration varies with the depth. This situation is the most interesting since it delivers adequate description of many surface physics problems. However, even in that case the DTM has been solved only numerically. Aers et al \cite{p8,p9} proposed a numerical algorithm for defect profiling using variable energy positron where defect concentration can vary with the depth like a step function. The drift of positrons in an external electric field has been included similarly. Van Veen et al.\cite{p10} presented the VEPFIT program which realizes the numerical approach to the DTM solution for the materials with layered structure. The program has frequently been used for treatment of the data obtained in the variable energy positron beam. This is because this program contains the procedure which allows to fit the different model parameters to the experimental data. Unfortunately, this program cannot be used for the pulsed beam technique for evaluation of  the measured positron lifetime spectra.

In the present paper, we report an exact analytical solution of the DTM for the case when a sample contains defects with inhomogeneous distribution.  Applying the Laplace transformation in time domain, we construct the Green's function of the time-dependent diffusion equation assuming that the defect depth profile is expressed by either the step function or a selected continuous function.
Our formalism permits to deduce the time evolution of positron fractions at different states and from this to obtain the PACh, including the positron lifetime spectrum. This was not possible until now with previously known DTM solutions. 

\section{The formalism}
\subsection {Statement of the problem}

In our consideration, a sample is a semi-infinite medium situated at positive semi-axis $0 \leq x< \infty$, Fig.1. Point $x=0$ plays a role of the medium surface, where the energetic positrons enter a sample. After a few picoseconds they thermalize and begin to diffuse. The time of diffusion beginning is considered as the initial time instant $t=0$. The penetration depth dependence of the initial positron concentration is called the positron implantation profile \cite{p11}; we denote it as $u(x,t=0)\equiv P(x)$. $P(x)$ is also a function of the incident  positrons energy. In our consideration, the total number of implanted positrons is normalized to unity:
\begin{equation}\label{eq0}
\int_0^\infty P(x)dx=1.
\end{equation} 

Generally speaking, a sample can contain defects with certain profile $C(x)$ close to its surface. The defect, like vacancy or their cluster can trap a positron with a specific trapping rate $\mu$. Thus the whole defect profile traps positrons at a rate $k(x)=\mu C(x)$, Fig.1. We denote the positron (which can freely diffuse) concentration in a bulk as $u(x,t)$. We denote the part of the above positrons trapped at the defects (vacancies) as $n_v(t)$. Certain positron fraction $n_{sur}(t)$ can diffuse back to the surface, which is a sink for thermalized positrons. We assume that the entering positrons cannot be trapped by the defects and the surface, i.e., $n_v(t=0)=n_{surf}(t=0)=0$. 

In the DTM model,  the concentration $u(x,t)$, obeys the diffusion equation which incorporates the annihilation rate in the free state $\lambda_{bulk}$ and spatially dependent trapping rate $k(x)$. This yields
 
\begin{equation}\label{fid1}
\frac{\partial u(x,t)}{\partial t}=D_+\frac{\partial^2u(x,t)}{\partial x^2}-[\lambda_{bulk}+k(x)]u(x,t),  
\end{equation}
where  $D_+$ is a bulk positron diffusion coefficient,  $0\le x<\infty$. As the thermalized positrons can be trapped at the surface with the absorption coefficient $\alpha$, the solution of the equation \eqref{fid1} should obey  the radiative boundary condition at the surface:
\begin{equation} \label{bk}
-D_+\left.\frac{\partial u(x,t)}{\partial x}\right|_{x=0}+\alpha u(x=0,t)=0.\\
\end{equation}

Positrons trapped at the surface can also annihilate with the rate $\lambda_{sur}$. This generates one more rate equation for the surface positron fraction $n_{sur}(t)$:
\begin{equation}\label{fid1a}
\frac{dn_{surf}(t)}{dt}=\alpha u(x=0,t)-\lambda_{surf}n_{surf}(t).
\end{equation}
Finally, the positrons fraction, trapped at the defects in a sample, obeys the equation:
\begin{equation}\label{fid1b}
\frac{dn_{v}(t)}{dt}=\int_0^\infty k(x)u(x,t)dx-\lambda_{v}n_{v}(t),
\end{equation}
where $\lambda_{v}$ is the annihilation rate at the defect. Subsequently we assume $\lambda_{bulk}>$ $\lambda_{surf}$ and $\lambda_{bulk}>$ $\lambda_{v}$. 

The set of equations \eqref{fid1} - \eqref{fid1b} shows that key function here is $u(x,t)$. Namely, if we know this function, the other equations can be solved easily as they do not contain spatial derivatives. The strategy 
of the solution of Eq. \eqref{fid1} is following. First, we fulfill Laplace transformation of time variable. Then, the resulting equation of the second order in $x$ should be solved with respect to the function $k(x)$. The analytical solution, however, is not possible for arbitrary $k(x)$. To show that, we can lower the order of the equation reducing it to Riccati form. It is well known (see, e.g. \cite{rwiki}) that Riccati equation cannot be solved analytically for arbitrary $k(x)$ function. However, for the specific forms of $k(x)$ the exact solutions can be found. 

Here we focus our attention on the simple "rectangular step" of the trapping rate in  the form:
\begin{equation}\label{fid3}
k(x)=\left\{\begin{array}{c}
k_0,\ x\leq x_0 \\0,\ x>x_0,
\end{array}\right.
\end{equation}
where $x_0$ is the range of defect profile. Such a profile has frequently been considered in the ion - implanted samples. Another profile, which is described by a smooth continuous function of the form
\begin{equation}\label{fid4}
k(x)=\frac{k_0}{1+e^{\gamma x}},
\end{equation}
also admits an analytical solution, but in the form of hypergeometic functions \cite{abr}, see below. The set of equations \eqref{fid1} -\eqref{fid1b} will be solved exactly with the above two defects profiles.

\begin{figure}
\begin{center}
\centering
\includegraphics[width=85mm]{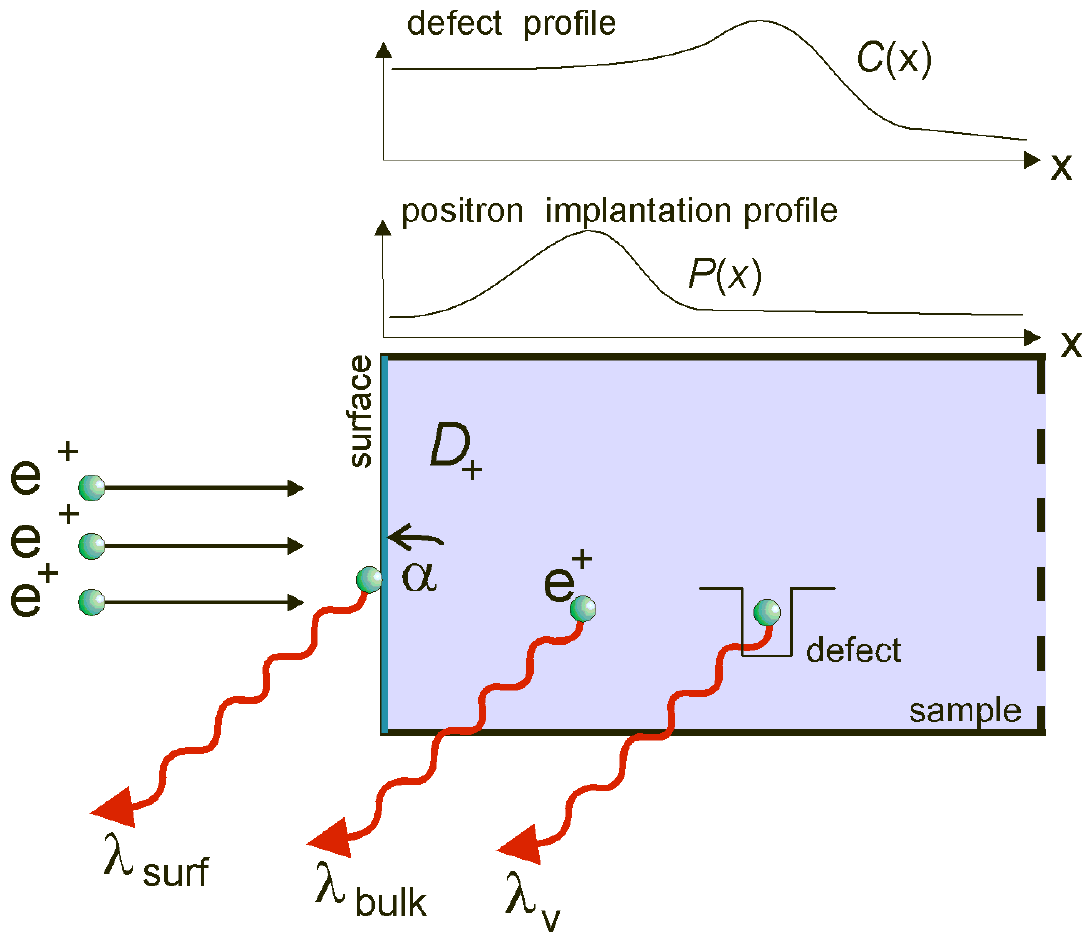} 
\caption{Sample irradiated by slow positrons and three states (considered in the paper) which can be occupied by them after thermalization (main lower panel). The positron annihilation occurs and the annihilation radiation is emitted from these states. The initial positron concentration or positron implantation profile $P(x)$ is presented in the middle panel. The defect depth profile, described by the $C(x)$ function and proportional to the trapping rate profile $k(x)=$ $\mu C(x)$, is reported in the upper panel.}
\end{center}
\end{figure}

\subsection{The method of problem solution.}
Here we discuss the method of exact analytical solution of the system \eqref{fid1}, \eqref{fid1a}, \eqref{fid1b} with respect to the boundary conditions \eqref{bk} and trapping rate profiles \eqref{fid3} and \eqref{fid4}. This solution, in turn, permits to calculate the experimentally measurable PACh.  In our approach, we first eliminate the time variable by Laplace transform in the eq.\eqref{fid1} as follows
\begin{equation}\label{lat}
\tilde{u}(x,s)=\int_0^\infty u(x,t)e^{-st}dt.
\end{equation}
As our problem has boundary \eqref{bk} and initial $P(x)$ conditions, the best way of its solution is Green's function method. The application of this method after elimination of the time variable in Eq. \eqref{fid1} gives

\begin{equation}\label{fid5a}
\tilde{u}(x,s)=\int_0^\infty P(\xi) G(x,\xi,s)d\xi,
\end{equation}
where $G(x,\xi,s)$ is Green's function obeying the boundary condition, i.e., the eq. \eqref{bk}. We will construct the Green's function from two fundamental linearly independent solutions of the corresponding {\em{homogeneous}} equation, which means that for our purposes the term $P(x)$ in Laplace transform can be omitted. The details of such procedure are contained in Appendix A. To find the solution of the eq.  \eqref{fid1} in time domain, the inverse Laplace transform is to be performed:

 \begin{equation}\label{fid5aa}
u(x,t)= L^{-1}\left[\int_0^\infty P(\xi) G(x,\xi,s)d\xi\right].
\end{equation}

For complete solution of the eqs. \eqref{fid1a}, \eqref{fid1b} in $s$ domain, we should have Laplace transformations of the fraction of positrons annihilating in a bulk
\begin{equation}\label{fr13}
\tilde{n}_{bulk}(s)=\int_0^\infty \tilde{u}(x,s)dx=\int_0^\infty dx \int_0^\infty P(\xi)G(x,\xi,s)d\xi
\end{equation}
at the surface
\begin{equation}\label{fr14}
\tilde{n}_{surf}(s)=\frac{\alpha}{s+\lambda_{surf}}\int_0^\infty P(\xi)G(x=0,\xi,s)d\xi
\end{equation}
and at the defects
\begin{equation}\label{fr15}
\tilde{n}_{v}(s)=\frac{1}{s+\lambda_{v}}\int_0^\infty k(x) dx \int_0^\infty P(\xi)G(x,\xi,s)d\xi.
\end{equation}
Two latter transformations are indeed the solutions of differential equations \eqref{fid1a}, \eqref{fid1b} in $s$ domain.

The above solutions permit to calculate the experimentally observable PACh. We begin with the positron lifetime spectrum, which is defined as  the probability of positron annihilation from any state in time interval ($t$, $t+dt$). The spectrum, normalized to the total number of implanted positrons, can be expressed as $-dn(t)/dt$, where the positron number $n(t)=n_{bulk}(t)+n_{surf}(t)+n_{v}(t)$. Integrating the equation \eqref{fid1} over $x$ (see also \eqref{fr13}) and adding the Eqs. \eqref{fid1a} and \eqref{fid1b}, we obtain 

\begin{equation}\label{fr16}
-\frac{dn(t)}{dt}=\lambda_{bulk}n_{bulk}(t)+\lambda_{surf}n_{surf}(t)+\lambda_{v}n_{v}(t).
\end{equation}

Expressions \eqref{fr13}, \eqref{fr14} and \eqref{fr15} permit to derive the corresponding PACh in time domain applying the inverse Laplace transform.
However, the another PACh, mean positron lifetime, can be calculated without explicit inversion of Laplace transform:
\begin{widetext}
\begin{equation}\label{fr17}
\bar{\tau}=\int_0^\infty t\left(-\frac{dn(t)}{dt}\right)dt=\int_0^\infty\left[n_{bulk}(t)+n_{surf}(t)+n_{v}(t)\right]dt=\tilde{n}_{bulk}(s=0)+\tilde{n}_{surf}(s=0)+\tilde{n}_{v}(s=0).
\end{equation}
\end{widetext}
Yet another important characteristic of positron annihilation lineshape is so-called  S - parameter. It is defined as the ratio of the number of counts in the line central part  to the total counts number under the annihilation line. This value can be obtained as follows:
\begin{widetext}
\begin{equation}\label{fr18}
S=S_{bulk}\lambda_{bulk}\tilde{n}_{bulk}(s=0)+S_{surf}\lambda_{surf}\tilde{n}_{surf}(s=0)+S_{v}\lambda_{v}\tilde{n}_{v}(s=0),
\end{equation}
\end{widetext}
where $S_{bulk}$, $S_{surf}$ and $S_{v}$ are S-parameters for positrons annihilating in the bulk, at the surface and on defects, respectively.

We can conclude that for the PACh calculations only two functions are needed, the initial implantation profile and the Green's function. The former profile, which depends on the positron implantation energy, can be obtained by Monte Carlo simulations performed, for instance, using the GEANT4 code. The analytical parametrization of this profile has been reported by several authors, the most popular is so-called Makhovian function. However, the Green's function should be derived by the explicit solution of the diffusion equation \eqref{fid1} for specific function $k(x)$. The procedure of such derivation accounts for boundary conditions and is described in many textbooks on boundary value problems for ordinary differential equations, see, e.g. \cite{cohi}. To make the paper self - contained, we briefly recapitulate this procedure in the Appendix A. This procedure will be used below to construct the Green's functions for the trapping rate profile expressed by eqs. \eqref{fid3} and \eqref{fid4}.
 
\section{DTM solution for the specific defect depth profiles}

\subsection {The homogeneous sample with uniformly distributed defects}

We begin with application of the above method to the solution of the problem with $k(x)=k_0=const$, which means that the defects are uniformly distributed across the sample.  After Laplace transformation the homogeneous part (which is the only needed for Green's function construction) of eq. \eqref{fid1} reads:
\begin{equation}\label{is}
D_+\frac{\partial^2 \tilde{u}(x,s)}{\partial x^2}- (\lambda_{bulk} +k_0+s)\tilde{u}(x,s)=0.
\end{equation}
The procedure of Appendix A permits to construct the Green's function for this case. While the explicit construction is given in Appendix B, here we present the final result in the form
\begin{widetext}
\begin{eqnarray}\label{fr11a}
&&{G}(x,\xi,s)=\frac{1}{\sqrt{4D_+s_1}}\Biggl\{\exp\left[-|x-\xi|\sqrt{\frac{s_1}{D_+}}\right]+\exp\left[-(x+\xi)\sqrt{\frac{s_1}{D_+}}\right]-\nonumber \\
&& -\frac{2\alpha}{D_+}\int_0^\infty \exp\left[-(x+\xi +\eta)\sqrt{\frac{s_1}{D_+}}-\eta \frac{\alpha}{D_+}\right]d\eta\Biggr\},\ s_1=s+\lambda_{bulk}+k_0.
\end{eqnarray}
This permits to obtain the explicit expressions for mean positron lifetime 
\begin{equation}\label{fr19}
\bar{\tau}=\frac{1}{\lambda_{surf}}+\frac{1}{\lambda_{bulk}+k_0}\left[1-\frac{\lambda_{bulk}}{\lambda_{surf}}+  k_0\left( \frac{1}{\lambda_{v}}-\frac{1}{\lambda_{surf}}\right)\right]\left[ 1- \int_0^\infty P(\xi)\frac{\exp\left( -\xi \sqrt{(\lambda_{bulk}+k_0)/D_+}\right)}{1+\sqrt{(\lambda_{bulk}+k_0)/D_+}/h}d\xi\right]
\end{equation}
and S - parameter ($h=\alpha/D_+$)
\begin{equation}\label{fr20}
S=\frac{S_vk_0+S_{bulk}\lambda_{bulk}}{\lambda_{bulk}+k_0}+\left( S_{surf}-\frac{S_vk_0+S_{bulk}\lambda_{bulk}}{\lambda_{bulk}+k_0} \right)\int_0^\infty P(\xi)\frac{\exp\left( -\xi \sqrt{(\lambda_{bulk}+k_0)/D_+}\right)}{1+\sqrt{(\lambda_{bulk}+k_0)/D_+}/h}d\xi
\end{equation} 
with the help of Eqs. \eqref{fr17} and \eqref{fr18} respectively.
\end{widetext}
For homogeneous sample, the calculations of the positron lifetime spectrum \eqref{fr16} can be performed on the base of Eq. \eqref{fr11a}. In this case, however, the inverse Laplace transform of the eqs. \eqref{fr13},  \eqref{fr14} and  \eqref{fr15} should be performed. The explicit  form of the inverse Laplace transform for Green's function \eqref{fr11a} is given by Eq. \eqref{fr12}. This generates 
following final expressions:
\begin{widetext}
\begin{eqnarray}
&&n_{bulk} (t) = \exp\left[-(\lambda_{bulk}+k_0)t\right] \times \nonumber \\
&&\times \int_0^\infty  P(\xi) \left[ {\mathrm{erf}\left( {\frac{\xi }{{\sqrt {4D_ +  t} }}} \right) + \exp\Bigl[h(\xi  + hD_ +  t)\Bigr]\mathrm{erfc}\left( {\frac{{\xi  + 2hD_ +  t}}{{\sqrt {4D_ +  t} }}} \right)} \right]d\xi, \label{fr21} \\ \nonumber \\
&& n_{surf} (t) = D_+  h^2 e^{- \lambda _{surf}\  t}\int_0^\infty  d\xi \ P(\xi)\int_0^t d\tau \ \exp \Bigl[ - (\lambda_{bulk}+k_0  - \lambda _{surf} )\tau \Bigr] \times \nonumber   \\
&&\times \left[\frac{1}{h\sqrt {\pi D_+ \tau }}\exp\left( - \frac{\xi ^2}{4D_+ \tau} \right) - \exp \Bigl[ h(\xi  + hD_ +  \tau ) \Bigr]\mathrm{erfc}\left(\frac{\xi  + 2hD_ +  \tau }{\sqrt {4D_ +  \tau }} \right) \right], \label{fr22} \\ \nonumber \\
&&n_v (t) = \frac{k_0\biggl[\exp \left[ - (\lambda_{bulk}+k_0) t \right] - \exp \left[ - \lambda _v t \right] \biggr]}{\lambda _v  - \lambda_{bulk}+k_0 }- \nonumber \\ 
&&- k_0 \exp ( - \lambda _v \,t)\int_0^\infty d\xi \, P(\xi)\int_0^t d\tau \exp \Bigl[ - (\lambda_{bulk}+k_0  - \lambda _v )\,\tau \Bigr] \times \nonumber \\
&&\times \left[ \mathrm{erfc}\left(\frac{\xi }{\sqrt {4D_ +  \tau }} \right) - \exp \Bigl[h\,(\xi  + hD_ +  \tau )\Bigr]
\mathrm{erfc}\left(\frac{\xi  + 2\,hD_ +  \tau }{\sqrt {4\,D_ +  \tau }} \right) \right].\label{fr23}
\end{eqnarray}
\end{widetext}
Here erf$(x)=(2/\sqrt{\pi})\int_0^\infty e^{-t^2}dt$ and erfc$(x)=1-$erf$(x)$ are error and complementary functions respectively \cite{abr}. 
Substitution of above expressions into Eq. \eqref{fr17} gives the formula for the desired positron lifetime spectrum. Note, that to use latter equation for experimental data treatment, we should make its convolution with corresponding time resolution function. Our calculations show that the case of homogeneous sample with uniformly distributed defects admits PACh to be expressed via quite simple expressions, which can be further easily calculated numerically. Assuming $k_0=0$ in the above relations yields the case of homogenous defect - free sample \cite{p7}. 

\subsection{Rectangular defect profile}
For the defect profile \eqref{fid3}, the homogeneous part of diffusion equation \eqref{fid1} after the Laplace transform renders as 

\begin{equation}\label{isx}
\left\{\begin{array}{c}
D_+\frac{\partial^2 u(x,s)}{\partial x^2}- (\lambda_{bulk} +k_0+s)u(x,s)=0,\ 0\leq x \leq x_0, \\
D_+\frac{\partial^2 u(x,s)}{\partial x^2}- (\lambda_{bulk}+s)u(x,s)=0,\ \ \ x>x_0.
\end{array}\right.
\end{equation}
The solutions of Eq. \eqref{isx}  read
\begin{eqnarray}\label{so1}
\left\{\begin{array}{c}u_1(x,s)=C_1e^{q_1x}+C_2e^{-q_1x},\ 0\leq x \leq x_0, \\
u_2(x,s)=C_3e^{q_2x}+C_4e^{-q_2x},\ \ \ \ \ \ \ x>x_0, 
\end{array}\right.\nonumber \\
q_1=\sqrt{\frac{\lambda_{bulk} +s +k_0}{D_+}},\ q_2=\sqrt{\frac{\lambda_{bulk} +s}{D_+}},
\end{eqnarray}
where $C_1$ - $C_4$ are arbitrary constants.
As we have outlined in Appendix A, the Green's function of the eq. \eqref{isx} should be constructed from the pair of linearly independent solutions for the entire real semi axis. These solutions, in turn, should be constructed from the functions \eqref{so1} so as to be continuous in the point $x_0$. To achieve that, we require that both the solutions $u_1$ and $u_2$ and their first derivatives should be equal in the point $x_0$.  
The next step is to construct the Green's function from the above two linearly independent and continuous at $x=x_0$ fundamental solutions. The continuity conditions at $x=x_0$ permit to express $C_1$ and $C_2$ via $C_3$ and $C_4$. The linear combination of obtained solutions generates following pair of functions
%\begin{widetext}
\begin{eqnarray}\label{shvv2}
u_1=\left\{\begin{array}{cc}
\cosh q_1(x-x_0)+q \sinh q_1(x-x_0), & x \leq x_0 \\
e^{q_2(x-x_0)}, & x>x_0,
\end{array}\right.\nonumber \\ \nonumber \\
u_2=\left\{\begin{array}{cc}
\cosh q_1(x-x_0)-q \sinh q_1(x-x_0), & x \leq x_0 \\
e^{-q_2(x-x_0)}, & x>x_0,
\end{array}\right.
\end{eqnarray}
%\end{widetext}
where $q={q_2}/{q_1}$. It is easy to check that functions $u_{1,2}(x)$ and their first derivatives are equal to each other at $x=x_0$. The procedure, outlined in Appendix A permits to construct the Green's function of the problem from the functions $Y_1(x,s)=C_1u_1(x,s)+C_2u_2(x,s)$ and $Y_2(x,s)=u_2(x,s)$. In this case, the boundary condition \eqref{bk} should be met by the function $Y_1(x)$ and namely its part for $x \leq x_0$.
This yields
\begin{eqnarray}
&&Y_1(x,s)=u_1(x,s)+\psi \ u_2(x,s), \label{u1} \\
&&Y_2(x,s)=u_2(x,s),\nonumber \\ \nonumber \\
&&\psi=-\frac{\left(q_1-\frac{\alpha}{D_+}q\right)\sinh x_0q_1+\left(\frac{\alpha}{D_+}-q_2\right)\cosh x_0q_1}{\left(q_1+\frac{\alpha}{D_+}q\right)\sinh x_0q_1+\left(q_2+\frac{\alpha}{D_+}\right)\cosh x_0q_1}.\nonumber 
\end{eqnarray}

The Greens function now has the form
\begin{equation}\label{gf1}
G(x,\xi)=\left\{\begin{array}{c} a(\xi)(u_1(x)+\psi u_2(x)), \ 0\leq x \leq \xi, \\
b(\xi)u_2(x), \ \ \  \xi \leq x < \infty.
\end{array}\right.
\end{equation}
The expressions for $a(\xi)$ and $b(\xi)$ can be found from Eq. \eqref{fg4}. They read
\begin{eqnarray}
&&a(\xi)=\frac{u_2(\xi)}{D_+\Delta(\xi)},\ b(\xi)=\frac{u_1(\xi)+\psi u_2(\xi)}{D_+\Delta(\xi)},\nonumber \\ 
&&\Delta(\xi)=u_2(\xi)u_1'(\xi)-u_1(\xi)u_2'(\xi)\equiv 2q_2. \label{gf2}
\end{eqnarray}
Explicitly
\begin{equation}\label{gf3}
G(x,\xi)=\frac{1}{2D_+q_2}\left\{\begin{array}{c} u_2(\xi)[u_1(x)+\psi u_2(x)], \ 0\leq x \leq \xi, \\
u_2(x)[u_1(\xi)+\psi u_2(\xi)], \ \ \  \xi \leq x < \infty.\end{array}\right.
\end{equation}
The Green's function \eqref{gf3} can be readily applied to the eqs. \eqref{fr13}, \eqref{fr14} and \eqref{fr15} to obtain the PACh calculated in the preceding subsection. In this case, the explicit expressions will be much more cumbersome then Eqs. \eqref{fr21} - \eqref{fr23}, however, the numerical calculations can be done quite easily. Even the positron lifetime spectrum can be recovered numerically.

\subsection{Continuous defect profile}
We now turn to the solution (which actually reduces to Green's function construction) of the DTM model for the case of continuous trappig rate profile \eqref{fid4}. Although in this case the solution is expressed via hypergeometric functions (see \cite{land3,abr}), it is useful for dealing with experimental situation as an example of analytically tractable continuous defect profile.

After Laplace transformation we obtain for homogeneous part of equation \eqref{fid1}
\begin{equation}\label{nfc1}
D_+\frac{d^2u(x,s)}{dx^2}-\left[\lambda_{bulk}+s+\frac{k_0}{1+e^{\gamma x}}\right]u(x,s)=0.
\end{equation}
Following Landau \cite{land3}, we make following change of variables in \eqref{nfc1}
\begin{equation}\label{nfc2}
\zeta=-e^{\gamma x},
\end{equation} 
which renders it to the form
\begin{equation}\label{nfc4}
\zeta^2(1-\zeta)u''+\zeta(1-\zeta)u'-\frac{u}{D_+\gamma^2}\left[s(1-\zeta)+k_0\right]=0.
\end{equation} 
Further substitution 
\begin{equation}\label{nfc5}
u(\zeta)=\zeta^{i\nu/\gamma}w(\zeta)
\end{equation} 
with so far unknown constant $\nu$ renders \eqref{nfc4} to the form, resembling the hypergeometric differential equation, see \cite{abr}. To reduce it exactly to hypergeometric form, the constant $\nu$ should be chosen as  
 \begin{equation}\label{nfc7}
 \nu^2=-\frac{\lambda_{bulk}+s+k_0}{D_+}.
 \end{equation}
The equation \eqref{nfc4} then assumes following hypergeometric form \cite{land3,abr}
\begin{equation}\label{nfc7a}
\zeta(1-\zeta)w''(\zeta)+[c-(a+b+1)\zeta]w'(\zeta)-ab\ w(\zeta)=0.
\end{equation}
The pair of fundamental, linearly independent solutions of Eq. \eqref{nfc7a} are hypergeometric functions \cite{land3,abr}
 \begin{eqnarray}
 w_1(\zeta)&=&F(a,b,c,\zeta),\label{nfc8} \\ 
 w_2(\zeta)&=&\zeta^{1-c}F(b-c+1,a-c+1,2-c,\zeta),\nonumber
 \end{eqnarray}
where the parameters are given by the expressions
\begin{eqnarray}
a&=&\frac{\sqrt{\lambda_{bulk}+s+k_0}-\sqrt{s}}{\gamma\sqrt{D_+}}, \label{nfc8a} \\
b&=&\frac{\sqrt{\lambda_{bulk}+s+k_0}+\sqrt{s}}{\gamma\sqrt{D_+}},\ c=1+a+b. \nonumber
\end{eqnarray} 
We note here, that the parameters of hypergeometric functions \eqref{nfc8a} turn out to be real.
Going back to variable $x$ by means of \eqref{nfc2} permits to obtain following explicit relations for 
fundamental solutions of the Eq. \eqref{nfc1}
\begin{widetext}
\begin{eqnarray}
&&u_1(x)=e^{\pi iQ }\exp\left[x\sqrt{\frac{\lambda_{bulk}+s+k_0}{D_+}}\right]
F[a,b,1+2Q,-e^{\gamma x}],\label{nfc12} \\
&&u_2(x)=e^{-\pi iQ }\exp\left[-x\sqrt{\frac{\lambda_{bulk}+s+k_0}{D_+}}\right]%
F[-a,-b,1-2Q,-e^{\gamma x}],\ Q=\frac{\sqrt{\lambda_{bulk}+s+k_0}}{\gamma\sqrt{D_+}}.\nonumber
\end{eqnarray}  
\end{widetext}
Note, that to pass from \eqref{nfc12} to the case considered in Ref. \cite{p7}, we should put $k_0 \to 0$ and $\gamma \to \infty$. In this case we have $a=0$ and $Q=0$ in \eqref{nfc12} and with respect to relation $F(a=0,b,c,z)=1$ \cite{abr}, we obtain from \eqref{nfc12} $u_{1,2f}(x)=e^{\pm x\sqrt{s_1/D_+}}$,which corresponds exactly to Eq. \eqref{fr3}. \\ \\

The boundary (at $x=0$) condition $u'_x-(\alpha/D_+)u=0$ in terms of variable $\zeta$ renders as
\begin{equation}\label{nfc13}
-\gamma u'_\zeta -\frac{\alpha}{D_+} u=0,\ \zeta=-1.
\end{equation}  
With respect to \eqref{nfc13}, the new set of fundamental solutions assumes the form
\begin{widetext}
\begin{eqnarray}
&&y_1(\zeta)=u_1(\zeta)+\psi\ u_2(\zeta),\nonumber \\
&&\psi=-e^{2\pi iQ}\frac{\left(\frac{\alpha}{D_+}-\gamma Q \right)F[a,b,1+2Q,-1]+\frac{k_0}{\gamma D_+}\frac{F[a+1,b+1,2(1+Q),-1]}{1+2Q}}{\left(\frac{\alpha}{D_+}+\gamma Q \right)F[-a,-b,1-2Q,-1]+\frac{k_0}{\gamma D_+}\frac{F[1-a,1-b,2(1-Q),-1]}{1-2Q}}.\label{nfc15b}
\end{eqnarray}
\end{widetext}
The above procedure permits to represent the Green's function in the form \eqref{fg3}, where $a(\xi)$, $b(\xi)$ and $\Delta(\xi)$ are given by \eqref{gf2} and $u_{1,2}$ by \eqref{nfc12}.
Final form of the Green's function yields
\begin{equation}\label{sol}
G(x,\xi,s)= \left\{\begin{array}{c}
-\frac{u_2(\xi)}{D_+ \Delta(\xi)}\left[u_1(x)+\psi u_2(x)\right] ,\ 0\leq x \leq \xi \\ \\
-\frac{u_1(\xi)+\psi u_2(\xi)}{D_+ \Delta(\xi)}u_2(x),\ \xi \leq x \leq \infty.
\end{array}\right.
\end{equation}
The equation \eqref{sol} is indeed an exact solution of the DTM model for the case of trapping rate profile \eqref{fid4}. Having the Green's function in $s-$ space, we can readily obtain the solution of our initial and boundary value problem using Eq. \eqref{fid5aa}. We note that the difficulties in numerical representation of hypergeometric function \cite{abr} make the explicit calculations to be hard in this case. In spite of this, the necessary APCh can still be obtained numerically. 

\section{Specific calculations and discussion of the theoretical results}

We test the obtained formulas for parameters corresponding to those for copper. The Monte Carlo (MC) simulations performed by Valkealathi and Nieminen \cite{Nieminen1, Nieminen2} and other authors had shown that the positron implantation profiles in semi-infinite materials can be best represented by the derivative of a Gaussian, which can be expressed in more general form by the so-called Makhovian function: 

\begin{equation}\label{t1}
P(x) = \frac{{m\,x^{m - 1} }}{{x_0^m }}\exp \left[ { - \left( {\frac{x}{{x_0 }}} \right)^m } \right],
\end{equation}
where $m$ and $x_0$ are adjustable parameters to fit the data. It is well established that only $x_0$ parameter depends on the incident positron energy $E$ (in keV) in the following manner:

\begin{equation}\label{t2}
x_0  = \frac{A_{1/2}}{\rho \,\ln(2)^{1/m}}E^n,
\end{equation}
where $\rho$ is the density of a host material, $A_{1/2}$ and $n$ are constants which depend on material. The MC simulation performed by GEANT4 codes shows that for Cu the parameters are equal to $m=1.729$, $n=1.667$ and $A_{1/2}=2.58$ $\mu$g/cm$^2$ so that $x_0=3.559E^{1.667}$, where $x_0$ is in nanometers and $E$ is in keV, \cite{p11}. Note, that in ref.  \cite{p11}, another function has also been suggested for description of the implantation profiles. However here we use the Makhovian function \eqref{t1}. We note here that $x_0$ in eq. \eqref{t1} should not be confused with the rectangular potential width in the eq. \eqref{fid3}. But as we actually use the energy dependence \eqref{t2} instead of $x_0$, this will not cause any misunderstanding.

The other DTM parameters are following. The annihilation rates $\lambda_{bulk}=1/117$ ps$^{-1}$, 
$\lambda_{surf}=1/450$ ps$^{-1}$. The positron diffusion coefficient $D_+=125.1$ nm$^2$/ps and surface trapping rate $\alpha=1.25$ nm/ps. 
In Fig. 2, we report the mean lifetime for both defect free sample (dashed line) and that with uniformly distributed defects (full line). Also,  the contributions from bulk $n_{bulk}$, surface $n_{surf}$ and defect (for a sample with uniformly distributed defects) $n_v$  annihilation channels are shown. All lifetimes are plotted against the incident positron energy and/or depth from the sample surface related by eq. \eqref{t2}. It is seen from Fig.2. that as the positron energy increases, the surface contribution decreases and bulk increases. This fact can be well understood since positrons implanted at larger depths have smaller chances to return back to the surface so that they annihilate mainly from the free state. The mean lifetime \eqref{fr19}, which for defect-free sample is a sum of above two contributions, also decreases with the increasing energy, and at energies above 50 keV the bulk value 117 ps is reached. We should emphasize that the dependencies reported in Fig. 2 could be extracted from the measurements using the pulsed beam technique. However, the popular VEP technique permits to extract (also as a function of the incident energy) only the linear combination of $\tilde{n}_{bulk}(s=0)$ and $\tilde{n}_{surf}(s=0)$ (eqs. \eqref{fr13}, \eqref{fr14}) as well as the S-parameter, eq. \eqref{fr20}, on their base. The decrease of S-parameter with the increase of the incident positron energy is observed for annealed metals. The typical length of such decrease is the positron diffusion length defined as $L_+=\sqrt{D_+\lambda_{bulk}}$. It can be extracted from the above experimentally observed decrease, which have been successfully corroborated in our previous papers. In semiconductors, like defect-free silicon, the opposite dependence is observed due to the fact that $S_{surf}<S_{bulk}$, the S-parameter increases with the increase of the incident positron energy, and at $E>30 keV$ the dependency saturates. This dependency can also be reproduced using the \eqref{fr20}. 

The full lines in Fig.2 show the case of a sample with uniform defect distribution having $\lambda_v=1/300$ ps$^{-1}$, $\mu=500$ ps$^{-1}$ and $C=2\cdot 10^{-5}$. This implies that the trapping rate equals to $k_0=0.01$ ps$^{-1}$. In this case, according to Eq. \eqref{fid1b}, the trapping at defects occurs. 
The defect contribution, $\tilde{n}_v(s=0)$, increases with the increase of the positron energy, however at energies above 30 keV it saturates. Due to the annihilation at defects the mean lifetime is increased as compared to the bulk value. Since the defects trap positrons so that not all of them return to the surface, the surface contribution is less than that in defect-free case. Such tendency (in the form of S-parameter dependency on the incident energy) can be observed for metals with defects or damaged surface, see ref. \cite{copper}. Note that in this case the mean lifetime saturates at $E>30 keV$ with saturation value 215.7 ps. This value can be obtained from the well known relation $\bar{\tau}=\frac{1}{\lambda_{bulk}}\frac{1+k_0/\lambda_v}{1+k_0/\lambda_{bulk}}$ valid within the standard trapping model. This shows that the DTM can be treated as the extension of this model. 

The next step is to consider a sample with rectangular defect profile. To be more specific, here we  assume that defects (which can capture positrons) are distributed only up to the depth of $x_0= 1000$ nm from the surface, they trap positron at a rate $k_0=0.01$ ps$^{-1}$. In Fig. 3, we show this profile as the grey shaded rectangle. According to Eq. \eqref{t2}, the defect profile extension 1000 nm corresponds to the positron energy 29.4 keV. In Fig. 3, we report our theoretical dependencies calculated using the eqs. \eqref{fr13}, \eqref{fr14} and \eqref{fr15} with respect to the corresponding Green's function,  eq. \eqref{gf1}.  Due to positron diffusion and convolution with the implantation profile $P(x)$, the  quantity $\tilde{n}_{v}(s=0)$ (i.e. defect contribution to the total positron lifetime) does not follow the rectangular shape of the defect profile. It has a maximum in the middle of the region where defects are extended, and then decreases. This is well seen as the bump (in the energy range 10 - 20 keV) in the energy dependence of the mean positron lifetime $\bar{\tau}$. Such a bump in the S-parameter has been observed by many authors in ion-implanted metal alloys, semiconductors and insulators \cite{si}. It has been related to the increase of the defect concentrations induced by the ions. In ref. \cite{copper}, the same peculiarity has also been observed in the copper sample, which surface was exposed to the dry sliding. Commonly, the simulation of such peculiarity can be well done with the VEPFIT program so that the rectangular defect profile can be modeled by that code. In Fig. 4, we compare the mean lifetimes at $x_0=$ 200, 600, 1000 and 1400 nm for the rectangular trapping rate profile \eqref{fid3}. It is well seen that for the lowest $x_0=200$ nm, the characteristic bump is small as the surface and defect contributions overlap. At higher depths the surface contribution decreases and the defect contribution is better recognized. Note that for small and large $x_0$ values the mean lifetime tends to the asymptotic cases of defect free sample and of the sample with uniform defect distribution respectively. These asymptotic cases are represented as the gray dashed lines in Fig. 4. 

For the continuous defect profile expressed by the eq. \eqref{fid4}, the similar dependencies can also be obtained. They are reported in Fig. 5. In this case, the Green's function \eqref{sol}, has been used.  We take following parameter values: $\gamma=0.03$ 1/nm and $k_0=0.01$ 1/ps. It can be seen from Fig.5, that continuous defect profile does not generate any additional peculiarities in the mean positron lifetime, as compared to those for rectangular one.

\begin{figure}
\begin{center}
\centering
\includegraphics[width=82mm]{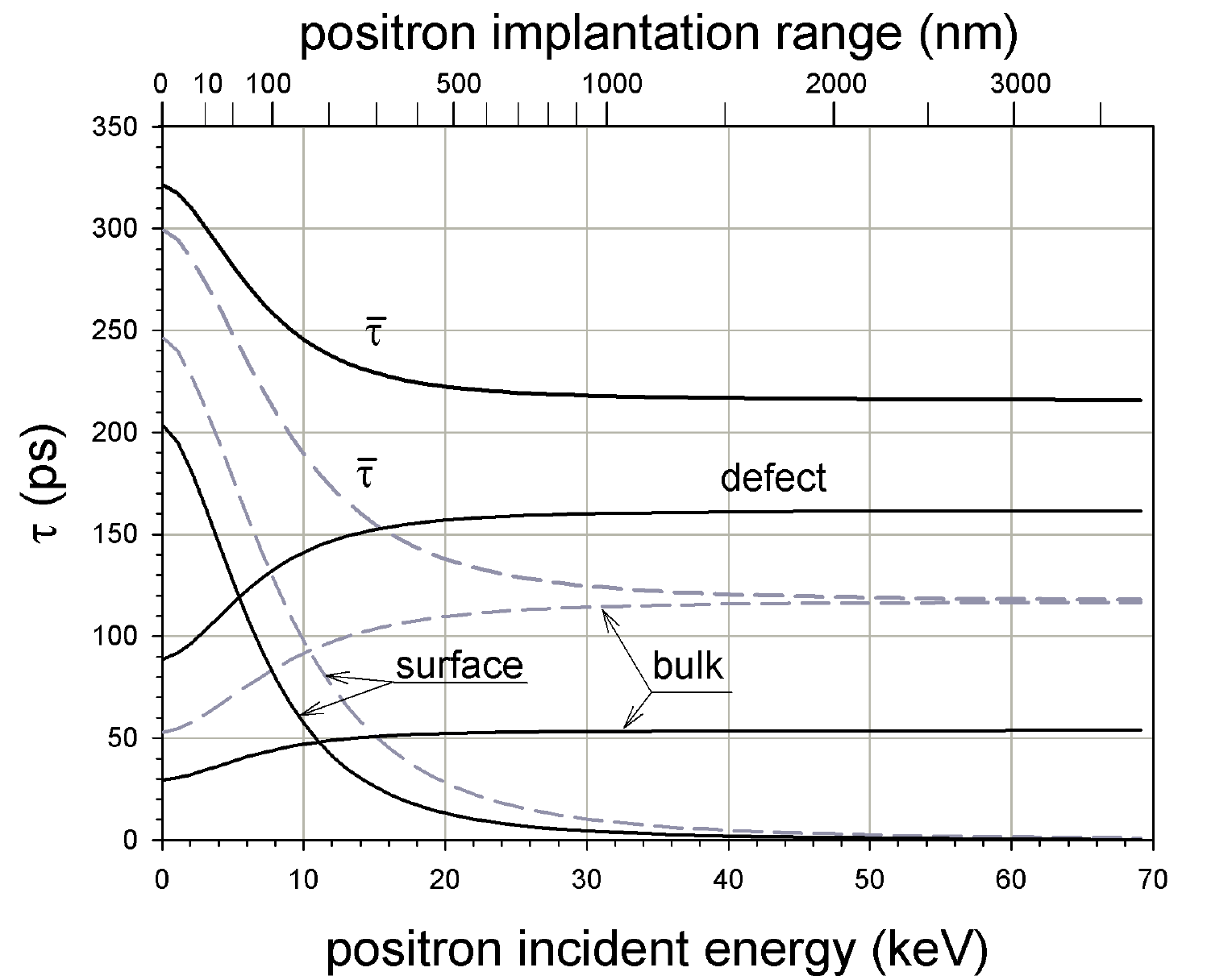} 
\caption{Total ($\bar{\tau}$) and partial ("defect", "surface" and "bulk") mean positron lifetimes as functions of positron implantation range and incident energy. Dashed lines correspond to defect free sample, solid gray lines to a sample with uniformly distributed defects. In the legend, 
"bulk" corresponds to $\tilde{n}_{bulk}(s=0)$, "surface" to $\tilde{n}_{surf}(s=0)$ and "defect" to $\tilde{n}_{v}(s=0)$.}
\end{center}
\end{figure}

\begin{figure}
\centering
\includegraphics[width=83mm]{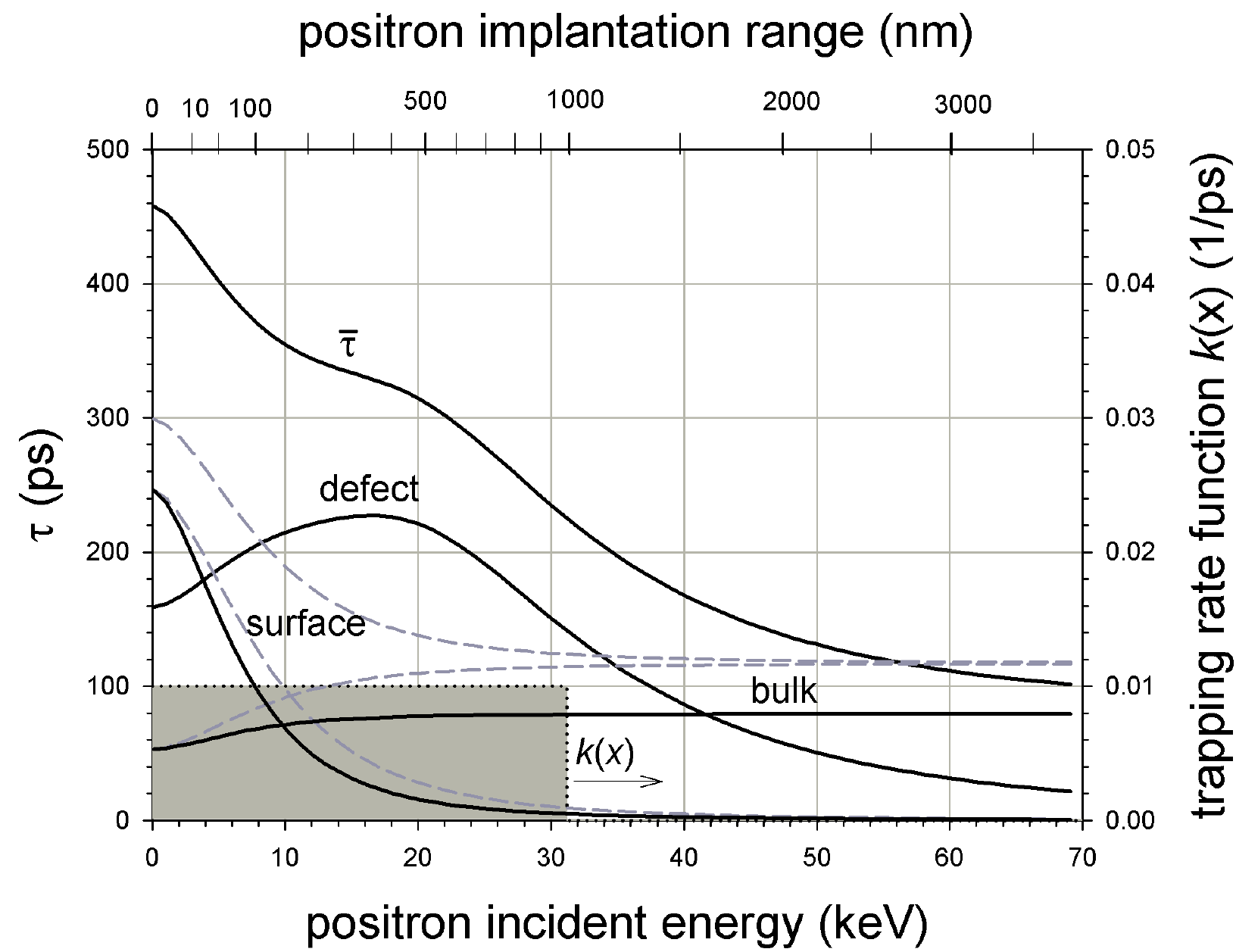}
\caption{Same as in Fig.2, but for rectangular defect profile \eqref{fid3} (shaded area) with the thickness $x_0=1000$ nm. The defects trap positrons at a rate 0.01 ps$^{-1}$. For comparison, the dashed gray line represents the defect free case taken from Fig.2.}
\end{figure}

\begin{figure}
\begin{center}
\centering
\includegraphics[width=83mm]{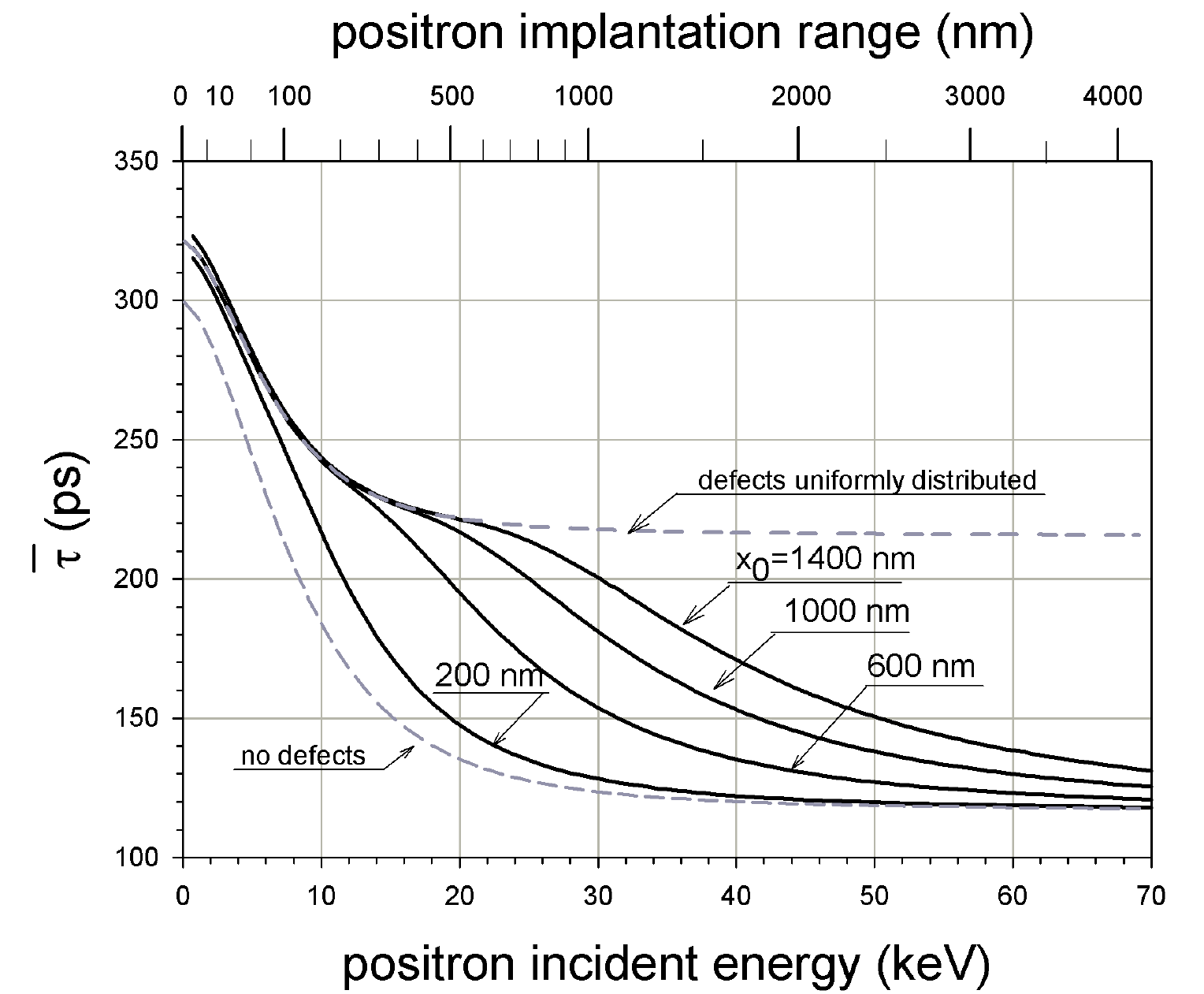}
\caption{Mean positron lifetime $\bar{\tau}$ as a function of positron implantation range and incident energy. The calculation corresponds to rectangular defect profile \eqref{fid3} for $x_0=$ 200, 600, 1000 and 1400 nm, shown in the panel. The assumed trapping rate is equal to  $k_0=0.01$ ps$^{-1}$. The upper and lower dashed gray lines correspond to the limiting cases from Fig.2.}
\end{center}
\end{figure}

\begin{figure}
\begin{center}
\centering
\includegraphics[width=83mm]{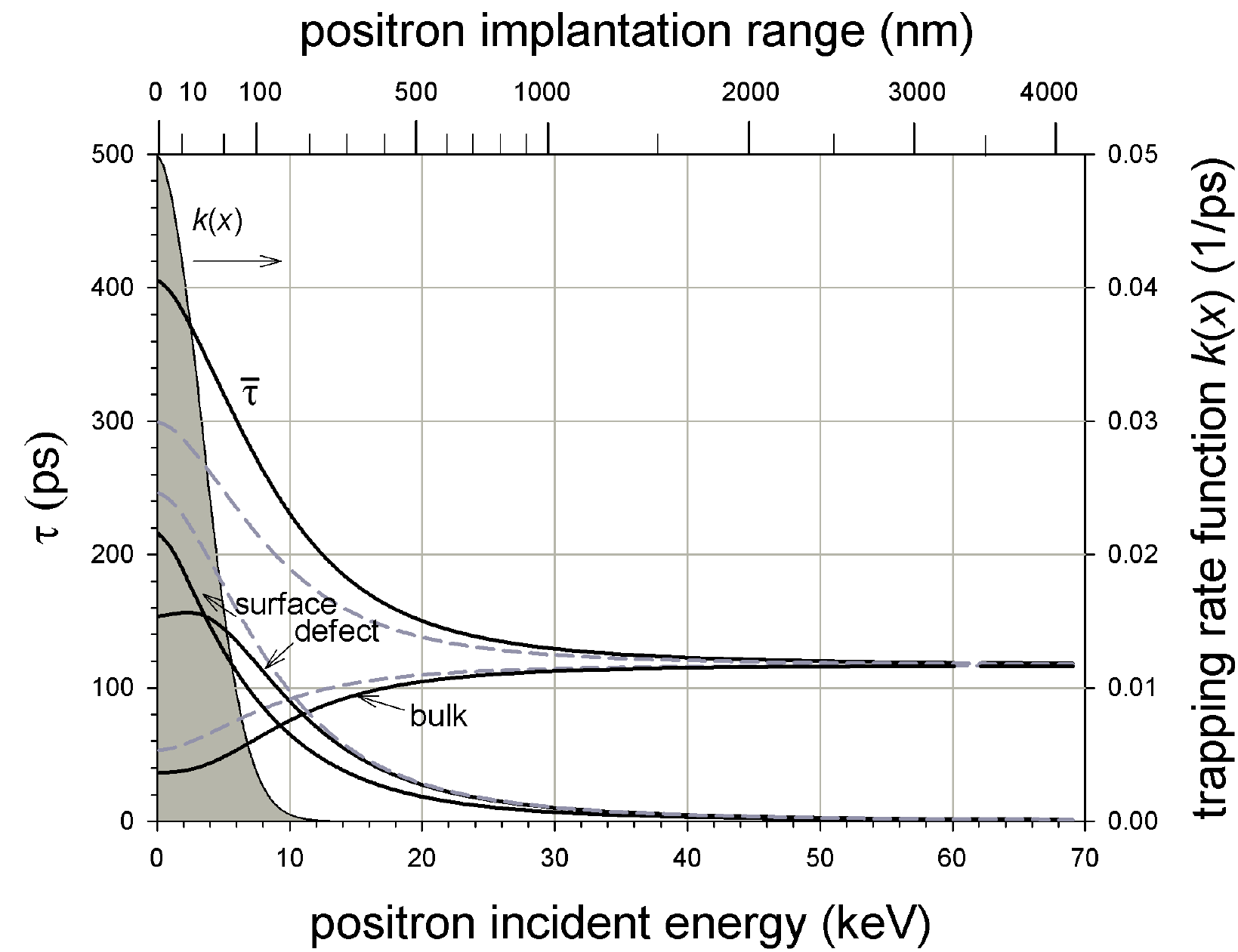}
\caption{Same as in Fig.2, but for continuous defect profile \eqref{fid4} (shaded area) with  $\gamma=0.03$ nm and $k_0=0.01$ ps$^{-1}$. For comparison, the dashed gray line represents the defect free case taken from Fig.2.}
\end{center}
\end{figure}

\section{Comparison with the VEPFIT code and experimental results }

One should note that in the real experiments the epithermal positrons also contribute to the measured S-parameter. According to Britton \cite{p5}, these positrons diffuse back to the surface and their fraction can be obtained as follows:

 \begin{equation}\label{exp1}
J_{epi}= \int_0^\infty P(x)\exp(-x/L_{epi})dx,
\end{equation}
where $L_{epi}$ is the scattering length, which value is around few nanometers. Since $P(x)$ depends on the positron incident energy, this fraction also exhibits the energy dependence. The measured value of the S-parameter now should be written as follows:

\begin{equation}\label{exp2}
S'=S(1-J)+S_{epi}J,
\end{equation}
where $S_{epi}$ represents the S-parameter corresponding to the epithermal positrons trapped at the surface. 
As it was mentioned above, the VEPFIT code is commonly used for description of the experimental dependencies of the S-parameter on the positron incident energy. The numerical algorithm in the code solves the steady - state diffusion equation for slow positrons in semiconductors with respect to defects and electric fields. The algorithm splits the sample into the intervals [$x$, $x+dx$] where the functions $P(x)$, $n_v(x)$ and electric field are constants. Then, in each of the intervals the solution of steady - state  eq. \eqref{fid1} has the form:
\begin{equation}
u(x)=A\exp{(\delta x)}+B\exp{(-\delta x)}+P/q,
\end{equation}
where $q=a^2/D_+(k n_v(x_i)+\lambda_{bulk})$, $\delta=\sqrt{\beta}$ and $a$, $A$, $B$ and  $P=P(x_i)$ are constants for each depth interval $x_i$. Applying the continuity conditions of $u(x)$ and $u'(x)$ at the interval boundaries, one can find $A$ and $B$ in the above intervals as the solution of the corresponding matrix equations. Additionally the sample is divided into layers of constant defect density. This allows to approximate the real defect depth distribution by the "stairs - like" function. The other details of the algorithm can be found in refs. \cite{p8,p9,p10}.

In Fig. 6, we present the results of the VEP experiment for defect - free Si and Si implanted with 300 keV Bi ions reported by Eichler et.al. \cite{si}. The open circles represent the data obtained for defect - free Si. The solid line represents the best fit of the equation eq. \eqref{fr20} with $k_0=0$ to these experimental points, while the dashed gray line represents the result of the VEPFIT fitting procedure. In both cases the Makhovian profile parameters have been taken from \cite{si}, namely $m=2$, $n=1.7$ and $A=$ 2.75 $\mu$g/cm$^2/$keV$^n$. Both theoretical dependences coincide very well with experimental points and with each other. The positron diffusion length (which is very important parameter, see above) extracted from both fits equals to $L_+=(202.5\pm 8.6)$ nm and $L_+=(203.9\pm 10)$ nm for VEPFIT and \eqref{fr20} with $k_0=0$, respectively. Two latter values are also in good coincidence. This shows that for the uniform, defect free case, the both (theoretical and numerical) approaches are equivalent. 
 
Let us consider the sample with defects. The full circles in Fig. 6 represent the experimental points for the Si sample were surface defect layer has been created by the implantation of the 300 keV Bi ions. This situation corresponds to the above case of rectangular defect profile. The data are well described by the VEPFIT program assuming two layers in a sample. The fitting result indicates that the upper layer thickness is about 358 nm and the diffusion length is about 61.2 $\pm$1.6 nm due to defects. The lower defect - free layer has larger diffusion length of 202 nm. The dashed gray line represents the VEPFIT fitting of the experimental points (full circles). The good coincidence is clearly seen. The solid black line shows the fit by our formulas for rectangular defect profile. According to our calculations, the layer with defects is about 700 nm thick and the trapping rate $k_0$ is about 0.045 1/ps. We can also observe the good coincidence between our theoretical results and experimental points. However, the small deviation from the VEPFIT results can be noticed. This deviation may be related to several factors. First of all the thickness of the adjoining surface layer with defects are about two times larger. Additionally we believe that the rectangular profile can be far from the real one. Also, the VEPFIT program simplifies the positron implantation profile $P(x)$, assuming it to be a constant in the each of the above intervals $x_i$. Our theoretical expressions permits to account accurately for the real (and arbitrary) shape of the profile $P(x)$. 

Fig. 7 reports one more comparison of the theoretical and numerical approaches to the experimental results treatment. This has been performed for pure copper sample exposed to the the dry sliding, reported in \cite{copper}. The gray dashed line depicts the VEPFIT fitting result indicating the presence of the defect layer close to the surface. The thickness of the layer is about 500 nm and the defect concentration has been estimated to be around $7\cdot10^{-5}$. The good agreement with the experiment is well visible. Similar agreement is achieved if we apply our formulas for rectangular defect profile with  $x_0=$ 500 nm and trapping rate around $4\cdot 10^{-4}$ ps$^{-1}$. The good agreement between the VEPFIT result and experimental points can be seen as well. 
 
 \section{Conclusions }
 
In this paper we present exact analytical solution of the trapping model which takes into account
the positron diffusion close to the sample surface irradiated by slow positrons. It has been assumed that thermalized positrons can annihilate at  the entering surface, in the host solid and at the defects distributed in the host. All these processes contribute to the measured signal, i.e. to the annihilation line shape parameter, lifetime spectrum or mean positron lifetime. 
The essence of our method is the Laplace transform in time domain with subsequent application of Green's function method to solve the resulting ordinary differential equation with initial and boundary conditions. The formalism is valid for virtually any defect profile $C(x)$, the details are presented in the Appendix.

The explicit expressions for Green's functions (and hence the problem solutions) have been derived for rectangular defect profile \eqref{fid3} and continuous one \eqref{fid4}. Probably, there exist other experimentally important profiles, for which the explicit analytical solutions can be obtained. 
 
We demonstrate that our general formalism and presented explicit solutions can be useful for profiling  the near - surface defect distributions by VEP technique. We have shown that our theoretical results are in pretty good agreement with numerical ones obtained by the VEPFIT program, which has been commonly used for the profiling. Moreover, our results allow to obtain the positron lifetime spectrum which could be measured using pulsed slow positron beam technique. Such calculation, however, requires to perform the inverse Laplace transform of our solutions. This problem is readily doable at least numerically.    
 
\begin{figure}
\centering
\includegraphics[width=83mm]{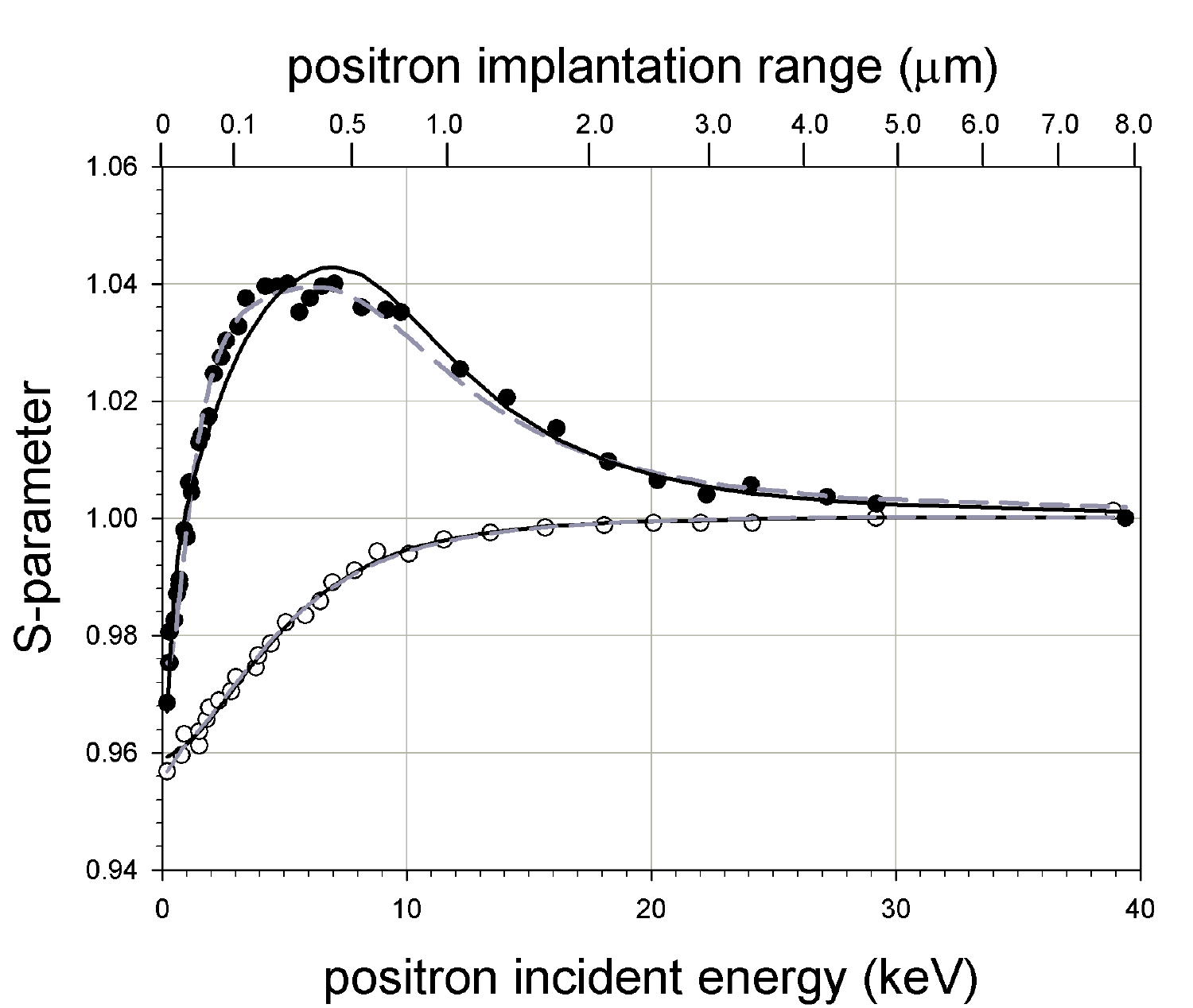}
\caption{Measured bulk S-parameter for annealed Si \cite{si} (open circles) and Si implanted 
with 300 keV Bi ions (full circles). Dashed gray lines represent the VEPFIT fitting of the experimental points. Black solid lines correspond to the calculation using eq. \eqref{fr18} for rectangular defect profile \eqref{fid3} with $x_0=300$ nm and $k_0=0.045$ ps$^{-1}$.}
\end{figure}

\begin{figure}
\centering
\includegraphics[width=83mm]{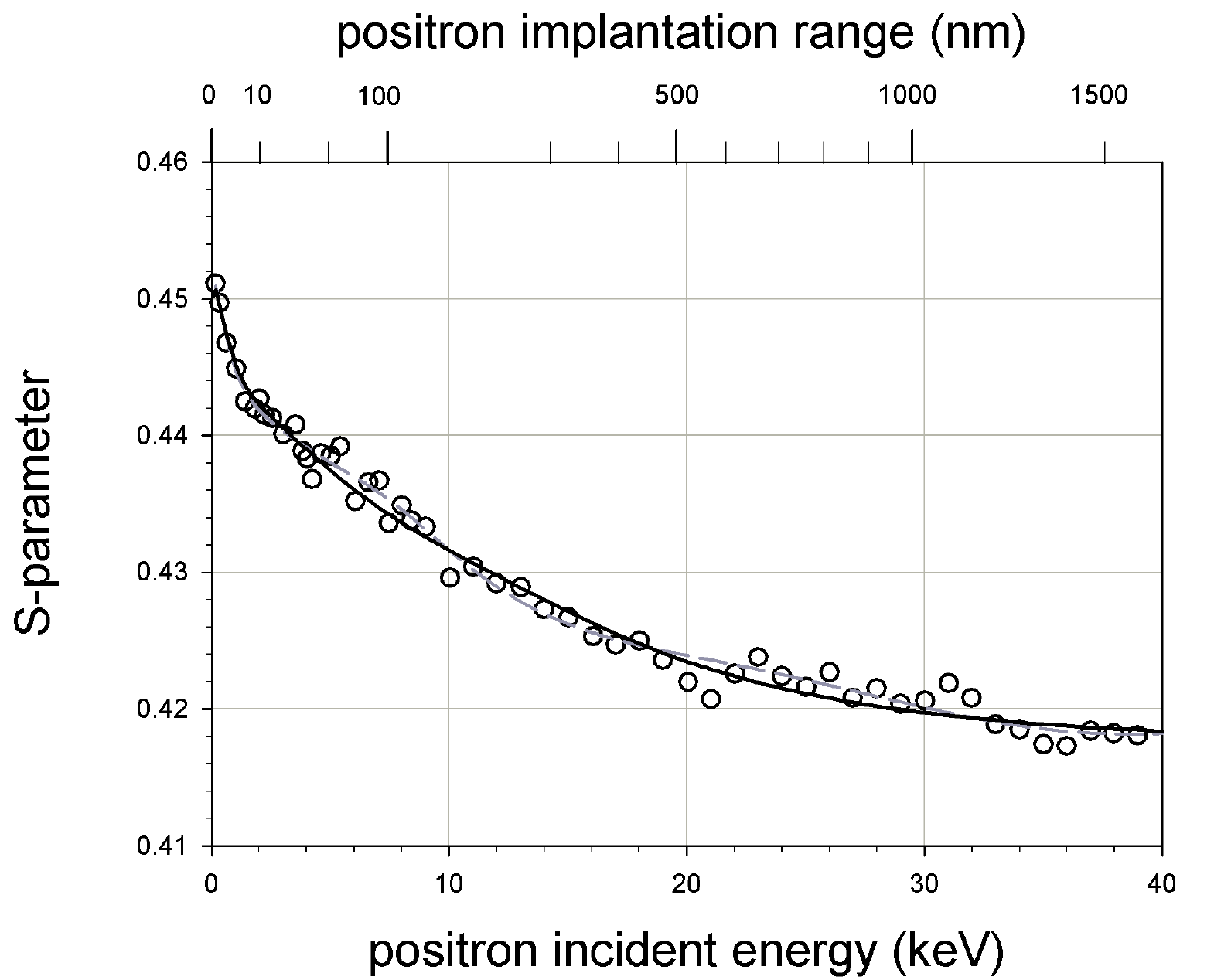} 
\caption{Measured S-parameter for Cu with surface being exposed to the dry sliding \cite{copper}.
Dashed gray lines represent the VEPFIT fitting of the experimental points. Black solid lines correspond to the calculation using eq. \eqref{fr18} for rectangular defect profile \eqref{fid3} with $x_0=$ 700 nm and $k_0=$0.067 ps$^{-1}$.}
\end{figure}

\appendix
\section{}

The Green's function can be constructed from two fundamental, linearly independent solutions of the {\em{homogeneous equation}}  i.e. that for $k(x)=0$ and $P(x)=0$. The procedure of such construction accounts for boundary conditions and is described in many textbooks on boundary value problems for ordinary differential equations, see, e.g. \cite{cohi}.
Generally, the Green's function of the second order ordinary differential equation 
\begin{equation}\label{fg1}
a_0(x)y''(x)+a_1(x)y'(x)+a_2(x)y(x)=0
\end{equation}
obeys the equation
\begin{equation}\label{fg2}
a_0(x)G''(x,\xi)+a_1(x)G'(x,\xi)+a_2(x)G(x,\xi)=\delta(x-\xi).
\end{equation}
At $x\neq \xi$ it obeys the equation \eqref{fg1} and at $x=0$ observes the boundary condition \eqref{fid1} (or any other). Now we assume that the equation \eqref{fg1} has $y_1(x)$ and $y_2(x)$ as the pair of its fundamental, linearly independent solutions. Then, Green's function $G(x,\xi)$ reads
  \begin{equation}\label{fg3}
  G(x,\xi)=\left\{\begin{array}{c}
  a(\xi)Y_1(x),\ 0 \leq x \leq \xi, \\ \\ b(\xi)Y_2(x),\ \xi \leq x \leq \infty, 
  \end{array}\right.
  \end{equation}
where $Y_1(x)=ay_1(x)+by_2(x)$ and $Y_2(x)=cy_1(x)+dy_2(x)$ should be chosen so that to satisfy the boundary condition \eqref{fid1}: $dY_1(x=0)/dx=(\alpha/D_+)Y_1(x=0)$. In our case we can safely assume $c=0$ and $d=1$ so that $Y_2(x)\equiv y_2(x)$. The functions $a(\xi)$ and $b(\xi)$ are determined by the conditions of continuity of Green's function and its derivative (actually the derivative has jump discontinuity) in the point $x=\xi$:
 \begin{equation}\label{fg4}
 a(\xi)Y_1(\xi)=b(\xi)Y_2(\xi),\ b(\xi)Y_2'(\xi)=a(\xi)Y_1'(\xi)+\frac{1}{a_0(\xi)}.
 \end{equation}
The application of this procedure gives the equations \eqref{gf1} and \eqref{sol} of the main text.

\section{}
The pair of fundamental solutions of the equation \eqref{is} has the form
\begin{equation}\label{fr3}
\tilde{u}_{1,2}(x,s_1)=\exp\left(\pm x\sqrt{\frac{s_1}{D_+}}\right),\ s_1=s+\lambda_{bulk}+k_0.
\end{equation}

According to procedure, outlined in Appendix A, we look for $Y_1(x,s_1)$ in the form $Y_1=Ae^{x\sqrt{s_1/D_+}}+Be^{-x\sqrt{s_1/D_+}}$ and substitute it to the boundary condition $Y_1'(x=0)-(\alpha/D_+)Y_1(x=0)=0$, which yields
\begin{eqnarray}
Y_1(x)&=&\exp\left(x\sqrt{\frac{s_1}{D_+}}\right)+\psi \exp\left(-x\sqrt{\frac{s_1}{D_+}}\right),\nonumber \\  Y_2(x)&=& \exp\left(-x\sqrt{\frac{s_1}{D_+}}\right),\ \psi=\frac{\sqrt{s_1D_+}-\alpha}{\sqrt{s_1D_+}+\alpha}.
\label{fr4a}
\end{eqnarray}
Here we choose $Y_2(x)=y_2(x)$ to have exponentially decaying solution at infinity. The Green's function now assumes the form
\begin{widetext}
\begin{equation}\label{fr5a}
\tilde{G}(x,\xi,s)=\left\{\begin{array}{c}
  a(\xi,s_1)\left[\exp\left(x\sqrt{\frac{s_1}{D_+}}\right)+\psi \exp\left(-x\sqrt{\frac{s_1}{D_+}}\right)\right],\ 0 \leq x \leq \xi, \\ \\ b(\xi,s_1)\exp\left(-x\sqrt{\frac{s_1}{D_+}}\right),\ \xi \leq x \leq \infty, 
  \end{array}\right.
\end{equation}
where 
\begin{equation}\label{fr7}
a(\xi,s_1)=-\frac{1}{2\sqrt{D_+s_1}}\exp\left(-\xi \sqrt{\frac{s_1}{D_+}}\right),\ b(\xi,s_1)=-\frac{1}{2\sqrt{D_+s_1}}\left[\exp\left(\xi\sqrt{\frac{s_1}{D_+}}\right)+\psi \exp\left(-\xi \sqrt{\frac{s_1}{D_+}}\right)\right].
\end{equation}
Substitution of explicit expressions \eqref{fr7} for coefficients into the expression for Green's function \eqref{fr5a}, permits to shorten it using modulus sign, namely

\begin{equation}\label{fr9}
\tilde{G}(x,\xi,s)=\frac{1}{\sqrt{4D_+s_1}}\left\{\exp\left[-|x-\xi|\sqrt{\frac{s_1}{D_+}}\right]+\psi \exp\left[-(x+\xi)\sqrt{\frac{s_1}{D_+}}\right]\right\},\ 0 \leq x \leq \infty.
\end{equation}
The representation of $\psi$ in the form
\begin{equation}\label{fr10}
\psi=\frac{\sqrt{s_1D_+}-\alpha}{\sqrt{s_1D_+}+\alpha} = 1-\frac{2\frac{\alpha}{D_+}}{\sqrt{\frac{s_1}{D_+}}+\frac{\alpha}{D_+}}=
1-2\frac{\alpha}{D_+}\int_0^\infty \exp\left[-\eta\left(\sqrt{\frac{s_1}{D_+}}+\frac{\alpha}{D_+}\right)\right]d\eta
\end{equation}
yields the known formula
\begin{equation}\label{fr9b}
\tilde{G}(x,\xi,s)=\frac{1}{\sqrt{4D_+s_1}}\left\{\exp\left[-|x-\xi|\sqrt{\frac{s_1}{D_+}}\right]+
\exp\left[-(x+\xi)\sqrt{\frac{s_1}{D_+}}\right]-\frac{2\alpha}{D_+}\frac{\exp\left[-(x+\xi)\sqrt{\frac{s_1}{D_+}}\right]}{\sqrt{\frac{s_1}{D_+}}+\frac{\alpha}{D_+}}\right\},
\end{equation}
which permits to obtain the equation \eqref{fr11a} of the main text from eq. \eqref{fr9}.
We perform the following inverse Laplace transformations \cite{abr}:
\begin{eqnarray}
&&L^{ - 1} \left[ {\frac{{\exp \left( { - \left| {x - \xi } \right|\sqrt {s/D_ +  } } \right)}}{{\sqrt {4D_ +  s} }}} \right] = \frac{1}{{\sqrt {4\pi D_ +  t} }}\exp \left[ { - \frac{{(x - \xi )^2 }}{{4D_ +  t}}} \right], \nonumber \\ 
&&L^{ - 1} \left[ {\frac{{\exp \left( { - (x + \xi )\sqrt {s/D_ +  } } \right)}}{{\sqrt {4D_ +  s} }}} \right] = \frac{1}{{\sqrt {4\pi D_ +  t} }}\exp \left[ { - \frac{{(x + \xi )^2 }}{{4D_ +  t}}} \right], \label{fr11} \\ 
&&L^{ - 1} \left[ {\frac{1}{{\sqrt {4D_ +  s} }}\frac{{\exp \left( { - (x + \xi )\sqrt {s/D_ +  } } \right)}}{{\sqrt {s/D_ +  }  + \alpha /D_ +  }}} \right] = \frac{1}{2}\exp \left[ {\alpha (x + \xi  + \alpha t)/D_ +  } \right]{\rm erfc}\left[ {\frac{{{\rm x + }\xi {\rm  + 2}\alpha t}}{{\sqrt {{\rm 4D}_{\rm  + } t} }}} \right].\nonumber
 \end{eqnarray}
 The substitution of \eqref{fr11} into \eqref{fr9} yields well-known form of the Green's function of the free diffusion problem on the semi-axis with mixed boundary condition:
 
\begin{eqnarray}\label{fr12}
G(x,\xi ,t) = \frac{\exp\left[\left(\lambda_{bulk}+k_0 \right)t\right]}{{\sqrt {4\pi D_ +  t} }}
\Biggl\{ \exp \left[ { - \frac{{(x - \xi )^2 }}{{4D_ +  t}}} \right] + \exp \left[ { - \frac{{(x + \xi )^2 }}{{4D_ +  t}}} \right] -\nonumber \\
- \frac{\alpha }{{D_ +  }} {\exp \left[ - {\alpha (x + \xi  + \alpha t )/D_+ }\right]}\ \mathrm{erfc}\left[\frac{x + \xi  + 2\alpha t }{\sqrt{4D_+t}} \right]  \Biggr\}. 
\end{eqnarray}
\end{widetext}

\end{document}